\documentclass[12pt,manuscript,center]{aastex}

\usepackage{epsfig}
\usepackage{bm}
\usepackage{subfigure}
\usepackage{multicol}
\usepackage{multirow}
\usepackage{array}
\usepackage{amsmath}
\usepackage{bm}
\usepackage{amsmath}
\usepackage{amssymb}
\usepackage{multirow}
\newcommand{\PreserveBackslash}[1]{\let\temp=\\#1\let\\=\temp}

\newcolumntype{C}[1]{>{\PreserveBackslash\centering}p{#1}}
\newcolumntype{R}[1]{>{\PreserveBackslash\raggedleft}p{#1}}
\newcolumntype{L}[1]{>{\PreserveBackslash\raggedright}p{#1}}

\shortauthors{Zhang et al.} \shorttitle{On estimating the force-freeness based on observed magnetograms}

\begin{document}

\title{On estimating the force-freeness based on observed magnetograms}

\author{X. M. Zhang\altaffilmark{1,2}, M. Zhang\altaffilmark{1} \& J. T. Su\altaffilmark{1}}

\altaffiltext{1}{Key Laboratory of Solar Activity, National Astronomical Observatories, Chinese Academy of Sciences, A20 Datun Road, Chaoyang District, Beijing 100012, China; \\ Email: xmzhang@nao.cas.cn}
\altaffiltext{2}{University of Chinese Academy of Sciences, China}

\begin{abstract}
It is a common practice in the solar physics community to test whether or not measured photospheric or chromospheric vector magnetograms are force-free, using the Maxwell stress as a measure. Some previous studies have suggested that magnetic fields of active regions in the solar chromosphere are close to be force-free whereas there is no consistency among previous studies on whether magnetic fields of active regions in the solar photosphere are force-free or not. Here we use three kinds of representative magnetic fields (analytical force-free solutions, modeled solar-like force-free fields and observed non-force-free fields) to discuss on how the measurement issues such as limited field of view, instrument sensitivity and measurement error could affect the estimation of force-freeness based on observed magnetograms. Unlike previous studies that focus on discussing the effect of limited field of view or instrument sensitivity, our calculation shows that just measurement error alone can significantly influence the results of force-freeness estimate, due to the fact that measurement errors in horizontal magnetic fields are usually ten times larger than that of the vertical fields. This property of measurement errors, interacting with the particular form of force-freeness estimate formula, would result in wrong judgments of the force-freeness: a truly force-free field may be mistakenly estimated as being non-force-free and a true non-force-free field may be estimated as being force-free. Our analysis calls for caution when interpreting the force-freeness estimates based on measured magnetograms, and also suggests that the true photospheric magnetic field may be further away from being force-free than they currently appear to be.
\end{abstract}


\section{Introduction}

It is well known that solar eruptive activities, such as filament eruptions, flares and coronal mass ejections, are closely related with the evolution of magnetic fields. That is, magnetic fields in the corona play a vital role in solar activities and give a way to understand the nature of solar eruptions (e.g. Zhang \& Low 2005). However, due to both intrinsic physical difficulties and observational limitations, direct measurement of the magnetic field in the corona is still difficult and only a limited number of examples are given (Cargill 2009; Lin et al. 2004). At present, an accepted way to overcome this difficulty is to reconstruct a coronal magnetic field  with the help of a force-free model, where the observed photospheric magnetic field is taken as a boundary condition (Wiegelmann \& Sakurai 2012; Wiegelmann et al. 2015). Under this approach, finding a force-free field suitable to be used as the boundary condition for extrapolation becomes important. As a first step, one therefore estimates whether an observed photospheric vector magnetogram is force-free or not.

Several previous studies have estimated the degree of force-freeness of the measured photospheric and chromospheric vector magnetic fields above active regions. Metcalf et al. (1995) estimated the force-freeness of active region NOAA 7216 and concluded that NOAA 7216 is not force-free \textbf{in} the photosphere but becomes force-free for heights beyond 400 km. This is the only work, to our knowledge, that measured the height-dependence of magnetic forces. 

Later on, Moon et al. (2002) analyzed 12 vector magnetograms of three flare-eruptive active regions (NOAA 5747, 6233 \& 6982) and concluded that the photospheric magnetic fields are not far away from being force-free. They also showed that the degree of force-freeness depends on the character of an active region and its evolutionary stage. Tiwari (2012) studied the high spatial resolution vector magnetic fields and also concluded that sunspot magnetic fields are not far away from being force-free even though their force-freeness may change with the time. However, Liu et al. (2013) carried out a statistical study of 925 active regions on the force-freeness of photospheric magnetic fields. They found that only about 25\% of the magnetograms can be considered as close to be force-free, i.e., most of the photospheric magnetic fields (75\%) are not force-free.

Whereas there is no consistency among previous studies on whether magnetic fields in the solar photosphere are force-free or not, different from previous studies, our study here is not to address whether a particular field is force-free or not, but to systematically study on how the limitations of magnetic field measurement (e.g. field of view, instrument sensitivity and measurement noise) could affect the judgment on force-freeness. We will use three kinds of magnetic field configurations as representative magnetic fields to show that the measurement limitations, the noise level in particular, will significantly influence the judgement of force-freeness and previous studies may have suffered from this limitation. The paper is organized as follows. We describe the method in Section 2. In Section 3 we analyze how the measurement limitations could affect the estimation of the force-freeness. Discussions are given in Section 4 and a brief summary is described in Section 5.


\section{The Method}

\subsection{The model}

Generally, it is assumed that the magnetic field is force-free in the corona (Wiegalmann et al. 2014), because model results suggest that the plasma $\beta$ (the ratio of the gas pressure to the magnetic pressure) is much less than unity ($\beta\ll1$) in the corona (Gary 2001). In this case, the magnetic field satisfies the following equations (see the reviews by Wiegelmann \& Sakurai 2012):
\begin{equation} \label{eq:eps}
\triangledown \times \mathbf{B}=\alpha ~ \mathbf{B} ~,
\end{equation}
\begin{equation}
\triangledown \cdot \mathbf{B}=0 ~,
\end{equation}
\noindent where {\bf B} is the vector magnetic field and $\alpha$ is the force-free coefficient.

Among all possible force-free fields, if  $\alpha$ = 0, then there is no electric current in the space and the field is a potential field; if $\alpha$ is a constant, then the field is called a linear force-free field; and in the real Sun, $\alpha$ usually varies spatially and the field is a general nonlinear force-free field (see Wiegalmann \& Sakurai 2012 for details).

According to Low (1985), in an isolated magnetic structure, a necessary condition for a force-free field to exist above a measured layer is:
\begin{equation}
F_x \ll F_p, \,\, F_y \ll F_p, \,\, F_z \ll F_p  ~,
\end{equation}
\noindent where $F_x$, $F_y$ and $F_z$ are the components of net Lorentz force and $F_p$  is the characteristic magnitude of the total Lorentz force which can be brought to bear on the atmosphere if the magnetic field is not force-free.
Assuming that the magnetic field above the plane $z = 0$ (photosphere) vanishes as z goes to infinity, the volume's boundaries other than the lower one do not contribute in the half-space $z>0$, so that the Maxwell stress can be written as following surface integrals:
\begin{equation}
\begin{aligned}
&F_x = -\frac{1}{4\pi} \int B_x\, B_z\, dx\,dy\,, \\
&F_y = -\frac{1}{4\pi} \int B_y\, B_z\, dx\,dy \,,\\
F_z &= -\frac{1}{8\pi} \int( B_z^2\,-B_x^2\,-B_y^2\,) \,dx\,dy\,, \\
F_p &= \frac{1}{8\pi} \int( B_z^2\,+B_x^2\,+B_y^2\,) \,dx\,dy \,,
\end{aligned}
\end{equation}
\noindent where $B_x$, $B_y$ and $B_z$ are the three components of the vector magnetic field, $B_z$ is the vertical magnetic field and
$B_x$, $B_y$ are the two components of the horizontal magnetic field $B_t$.

Above approach to estimate the force-freeness using measured vector magnetograms as the boundary condition at $z=0$ plane has become a common practice in the community. Whereas for an entirely force-free magnetic field, all the $F_x$, $F_y$ and $F_z$ must be zero, a magnetic field may be considered as being nearly force-free if the magnitudes of $F_x/F_p$, $F_y/F_p$, and $F_z/F_p$ are sufficient small (Low 1985). Metcalf et al. (1995) suggested a value of $|F_x/F_p|$, $|F_y/F_p|$, and $|F_z/F_p|$ less than 0.1 as a criteria for a measured magnetic field to be considered as being force-free. Following Metcalf et al. (1995) and others (e.g. Moon et al. 2002; Tiwari (2012); Liu et al. 2013), we will adopt this criteria in the following.

\subsection{Representative magnetic fields}

We will use three magnetic fields as representative fields to discuss the possible influences of measurement limitations on the force-freeness estimate.

The first one is taken from analytical solutions provided in Low \& Lou (1990). Under a few assumptions, such as the fields being axis-symmetry and the solutions separable in r and $\theta$ directions, Low \& Lou (1990) reformulated Equations (1) and (2) into a second-order partial differential equation in spherical coordinates with constant $n$ and $m$, where integer $n$ describes the power-law decreasing index in r-direction and $m$ defines the number of node points in $\theta$-direction. Finding eigenvalue solutions of this second-order partial differential equation generates a series of analytical nonlinear force-free fields. These fields can be transformed into Cartesian coordinates by arbitrarily positioning a plane, characterized by the parameters $l$ and $\varPhi$, to produce 3D force-free magnetic fields that represent the magnetic fields over active regions with a striking geometric realism. Here $l$ is the distance between the surface plane and the point source location, and $\varPhi$ is the angle between the axis of symmetry of the magnetic field and the z-axis in the Cartesian coordinate system. These analytical force-free solutions are quite useful in testing properties of force-free fields. For example, they have been extensively used to test the reliability and accuracy of force-free field extrapolation algorithms (e.g. Schrijver et al. 2006).

We use one of these analytical force-free solutions as a representative field. The field is generated by $n=1, \,m=1,\, l=0.3 $ and $\varPhi = \pi/2$ and is shown in Figure 1$a$. Here the contours show the vertical field strength with solid contours indicating $B_z>0$ field and dashed contours presenting $B_z<0$ field. The contour interval is 180 G. Green squares outline the ranges of different field of views (FOVs) to simulate different sizes of FOVs in real observations. The smaller the red labeled number is, the larger the FOV is.

To make the field more comparable to real magnetograms, we have re-scaled the field to make the maximum of $|B_z|$ be 2000 G.
The number of 2000 G is not specifically chosen among the possible values from 1000 G to 5000 G of active regions. The re-scaling is done merely to make our studied fields comparable. The particular number chosen would not influence our results qualitatively because the estimate of the force-freeness is based on the ratio, not the absolute values, of $F_x$, $F_y$ or $F_z$, to $F_p$. Some properties of the field are listed in Table 1.

\begin{table}[htbp]
\centering
\centerline{\footnotesize \textbf{Table 1:} Information of the three representative magnetic fields}
\label{tab1}
{\scriptsize
\begin{tabular}{cccccccc}
\hline
\hline
 & & &\multicolumn{3}{c}{FOV Number = 0} &  \\
 \cline{4-6}
 & Date \& Time (UT) & Position & $F_x/F_p$ & $F_y/F_p$ & $F_z/F_p$ & $max(B_z) (G)$ & $min(B_z) (G)$ \\
\hline
Low \& Lou & --- & --- & 1.57$\times10^{-8}$ & -0.00036 & -0.00468 & 2000.00 & -699.31 \\
AR11072 & 2010 May 23: 0500 & S14W00 & 0.00149 & 0.00169 & -0.02141 & 2000.00 & -1130.34 \\
AR10960 & 2007 June 07: 0304 & S07W07 & 0.07035 & -0.03191 & -0.51746 & 1506.42 & -2000.00\\
\hline
\end{tabular}
}
\end{table}

The second representative field is a modeled solar-like nonlinear force-free field that might exist on the Sun. In other words, we treat this modeled field as a more or less theoretical field but have a geometry more similar to a realistic one on the Sun, if exists, than the analytical one given by Low \& Lou (1990). We use a vector magnetogram of NOAA 11072 obtained by HMI/SDO (Schou et al. 2012) as the $z=0$ boundary and carry out a 3D force-free field extrapolation. We select a time when the active region is near the disk center, at a position of S14W00. We use the magnetogram from the hmi.sharp\_cea\_720s series, which has been inverted by the HMI team using the Milne-Eddington (ME) inversion algorithm of Borrero et al. (2011), solved the 180$^\circ$ ambiguity using the minimum energy method (Metcalf 1994) and processed using a cylindrical equal area (CEA) projection. We then further preprocess the magnetogram using methods described in Wiegelmann et al. (2006, 2008) to make it suitable for force-free extrapolation. The extrapolation is done with the help of an optimization code described in Wiegelmann (2004).

We use the layer about 1 Mm up than the $z=0$ photosphere in the extrapolated field as the second representative field: the modeled solar-like nonlinear force-free field. Note that 1 Mm up than the $z=0$ photosphere makes this field already into the layer of chromosphere. However, our purpose here is not to discuss what the true chromospheric magnetic field might be, but to get a modeled solar-like force-free field. So the particular height, 1 Mm or even 2 Mm up, make no difference for the results of our study here. In addition, to make all our representative fields comparable, the maximum value of $|B_z|$ of this field has also been normalized to be 2000 G. Some properties of this field are also shown in Table 1.

To quantify how good our force-free extrapolation is, we have calculated a few numbers as those in DeRosa et al. (2015). They are: $<CWsin\theta>=0.37$ and $<|f_i|> = 7.9\times10^{-4}$. These numbers are of the same magnitudes as those in DeRosa et al. (2015). In particular, it shows the success of the force-free extrapolation when we get very small numbers of $F_x/F_p=0.00149$, $F_y/F_p=0.00169$ and $F_z/F_p=-0.02141$ of the modeled solar-like field. The original observed magnetogram has these numbers as $F_x/F_p=-0.00207$, $F_y/F_p=0.08495$ and $F_z/F_p=-0.03457$. We see that the $F_y/F_p$ value has reduced to be 2\% of its original one.

Figures 1$b$ shows the $B_z$ map of this modeled solar-like force-free field. As before, colored rectangles here in Figure 1$b$, labeled with sequential numbers, outline ten different FOVs, to mimic limited FOVs in real observations. Again, the smaller the labeled number is, the larger the FOV is.

The third representative field is a non-force-free field. According to Tiwari (2012), the fields of active region NOAA 10960 are non-force-free in that $F_x/F_p=0.137$, $F_y/F_p=0.093$ and $F_z/F_p=-0.482$. We used a single vector magnetogram of this active region, obtained by the Solar Optical Telescope$/$Spectro-Polarimeter (SP) on Hinode (Kosugi et al. 2007). The SP magnetograms are inverted by the SP team from Stokes profiles using the MERLIN ME inversion algorithm (Skumanich \& Lites 1987; Lites et al. 2007) and the inherent 180$^\circ$ azimuth ambiguity is resolved in the same way as that for HMI data. Again, we choose a time, 03:04 UT on 2007 June 7, when the active region is near disk center. Also, the maximum of $|B_z|$ is  rescaled to be 2000 G, to make the three representative fields comparable to each other. Information of this field can be found in Table 1 and the B$_z$ map of this field is shown in Figure 1$c$ with the colored rectangles again representing different FOVs.

\subsection{Mimic the effect of FOV}

As already stated in Canfield et al. (1991), a limited FOV may not be appropriate for Maxwell stress integration, using which the flux balance is a pre-requisite for. To minimize this effect, Moon et al. (2002) only considered magnetograms whose magnetic imbalance (MI) is within 10\%. Metcalf et al. (1995) as well as Tiwari (2012) used this approach too. However, the effect of limited FOV on the force-free measure has not been studied systematically. Aiming to discuss on how the FOV could influence the judgment of the force-freeness, we shrink the FOV from the original flux-balanced one to get a series of magnetograms with different FOVs, as indicated by the colored rectangles in Figure 1.

A discussion on the effect of a limited FOV is actually related with the effect of flux imbalance (MI). We follow Moon et al. (2002) to calculate the MI associated with different FOVs as the index to represent different FOVs. MI is defined as:
\begin{equation}
MI = \frac{|F^+ - F^-|}{F^+ + F^-} \times 100  ~~,
\end{equation}
\noindent where $F^+ $and $F^-$ are upward ($B_z>0$) and downward ($B_z<0$) magnetic fluxes respectively.

\subsection{Mimic the effects of sensitivity \& noise}

Observed magnetograms also suffer from issues of instrument sensitivity and measurement noise. The random errors (noises) in the horizontal field measurement are particularly large, typically ten times of the vertical one. For example, random errors in HMI are about 5 G in the line-of-sight component whereas the uncertainty in the transverse field is between 70G and 200G (Wiegelmann et al. 2012; Hoeksema et al. 2014).

To deal with this situation, in previous studies usually only data points whose field strengths are larger than a certain value are used. For example, Metcalf et al. (1995) only use data points with magnetic field strength greater than 150 G (1 $\sigma$ noise level in transverse magnetic field). Moon et al. (2002) used field strength larger than 100 G as a criteria. Liu et al. (2013) used data points whose $|B_z|>20 G, |B_x| >150 G$ and $|B_y| >150 G$. However, this `cutting' method is more equivalent to setting a low level of sensitivity, the large measurement error (noise) is still buried in the remaining data points.

We mimic the sensitivity and noise separately to divide the effects of these two issues. Firstly, to simulate the different levels of sensitivity, set $B_t^0$ , $B_z^0$ as the horizontal and vertical field sensitivity respectively. If $ \sqrt{B_x^2+B_y^2}\leq  B_t^0$ or  $|B_z|\leq  {B_z^0}$, then omit these pixels.
Taking knowledge from previous studies, we have assumed $B_x^0$ and $B_y^0$ = 10 $B_z^0$ and hence $B_t^0 = 10\sqrt{2} B_z^0$. We have studied for $B_z^0$ increasing from 0 G, with a constant step size of 1 G. Only the magnetograms with the largest FOV is studied. To quantify this undertaking, a number NP, defined as the percentage of the data points that have been omitted, is calculated for each $B_z^0$ level.

Similarly, given $\sigma_x$, $\sigma_y$, $\sigma_z$ as the white noise in $B_x$, $B_y$ and $B_z$ respectively, to mimic the different levels of noise in the observed magnetograms, we have replaced the value of each pixel in the magnetogram using following method: $B_x$ is placed by $B_x + \sigma_x$, $B_y$ is placed by $B_y + \sigma_y$ and $B_z$ is placed by $B_z + \sigma_z$. $\sigma_z$ is created by $\sigma_z^0$ multiplying a normally-distributed random numbers, similarly for $\sigma_y$ from $\sigma_y^0$ and $\sigma_x$ from $\sigma_x^0$ . Again, we have assumed $\sigma_x^0$ and $\sigma_y^0$ = 10 $\sigma_z^0$, $\sigma_z^0$ increasing from 0 G with a constant step of 1 G.

\section{Analysis and results}

In this section, we quantify how different field of view (Section 3.1), instrument sensitivity (Section 3.2) and measurement noise (Section 3.3) could affect the estimation of force-freeness for analytical force-free solutions, extrapolated force-free fields and observed non-force-free fields, respectively.

\subsection{The influence of field of view}

Figures 2 shows how the different sizes of FOVs could influence the estimation of the force-freeness. Plotted here are the values of $F_x/F_p$ (blue lines), $F_y/F_p$ (red lines) and $F_z/F_p$ (black lines), obtained by Equation (4), as a function of the FOV number. Meanwhile, the variation of corresponding MI with each FOV, calculated by Equation (5), is also shown but in the right panels.

Panel $a$ shows the results for the analytical force-free field. It shows that with the increase of the FOV number (that is, with the decrease of the size of the FOV) the values of $F_x/F_p$, $F_y/F_p$, and $F_z/F_p$ can increase from the theoretical zero of force-free fields (FOV Number 0) to a magnitude larger than 0.1 (FOV Number 9). This suggests that a true force-free field (where $F_x/F_p=0$, $F_y/F_p=0$ and $F_z/F_p=0$) may be mistakenly estimated as being non-force-free (where $|F_x/F_p|>0.1$, $|F_y/F_p|>0.1$ or $|F_z/F_p|>0.1$) at FOV Number 9, where magnetic imbalance is larger than 90\% (see Panel $b$). However, before FOV Number 7, the changes of $F_x/F_p$, $F_y/F_p$, and $F_z/F_p$ are all small (within a magnitude of 0.1). This suggests that the wrong judgment may not happen even for a MI value as large as 43\% (FOV Number 7) and a criteria setting as MI less than 10\% is pretty safe.

As in Panels $a$ and $b$, Panels $c$ and $d$ show the results for the NOAA 11072 magnetogram based modeled solar-like force-free field. Before FOV Number 8, the changes of $F_x/F_p$, $F_y/F_p$ and $F_z/F_p$ are small and within 0.1, again suggesting that setting the MI within 10\% is safe. Also, if the FOV is too small where the MI is too large, a wrong judgment can be made, as is the case for FOV number 8. Interesting is that, under some circumstances, such as for the FOV Number 9, the severe flux imbalance (MI $\sim$ 40\%) does not seem influencing the measure, which tells that the MI is not the only factor that controls the force-freeness estimate.

 Panels $e$ and $f$ are for the observed non-force-free fields (NOAA 10960). This field shows a large MI variation with the increase of FOV number. However, a fair judgment can still be made if we confine the MI within 10\%, as before FOV Number 3 for this field. So, again we see 10\% of MI as a good criteria.

In summary, for the three representative fields, we see that a limited FOV with a large MI value indeed can bring certain effects on the results of force-freeness estimation as previous studies have suggested (Canfield et al. 1991, Moon et al. 2002), and by setting a criteria such as using magnetograms whose MI is within 10\% as suggested by many previous studies is a safe approach.

\subsection{The influence of instrument sensitivity}

Figure 3 presents the results of the influence of instrument sensitivity on force-freeness measurement. Variations of the $F_x/F_p$, $F_y/F_p$ and $F_z/F_p$ are shown in left panels. The right panels show the variations of MI (lines with cross symbols) and of NP (lines with triangle symbols). The lower x-axis is the $B_z^0$ in the unit of Gauss and the upper x-axis is the $B_x^0$ or $B_y^0$ also in the unit of Gauss.

Again Panels $a$ and $b$ are for the analytical force-free field, and Panels $c$ and $d$ are for the modeled solar-like force-free field. Here we see that the variations of $F_x/F_p$, $F_y/F_p$ and $F_z/F_p$ are all with 0.1 in magnitudes for these two force-free fields. This suggests that the problem of instrument sensitivity does not influence the force-freeness estimation seriously, at least for these two cases. For the analytical force-free field, even when MI has reached to 52\% and even when 90\% data points have been omitted ($B_z^0$ = 25 G case in Figure 3$b$), the change of $F_x/F_p$, $F_y/F_p$ and $F_z/F_p$ are all within 0.05 in magnitudes. The same is true for the modeled solar-like force-free field, even when MI has increased to 30\% and more than 90\% data points have been omitted ($B_z^0$ = 25 G in Figure 3$d$), the change of $F_x/F_p$, $F_y/F_p$ and $F_z/F_p$ are all within 0.06 in magnitudes..

For the observed non-force-free field as shown in Panels $e$ and $f$, however, the problem of instrument sensitivity would not obviously influence the force-free measures  only if $B_z^0$ less than 12 G or MI is less than 10\%. Therefore, we see that the instrument sensitivity would not influence the force-freeness estimates too seriously as long as MI is controlled within the safe criteria of 10\%.

\subsection{The influence of measurement noise}

Figure 4 shows the effect of measurement noise for the three representative fields, again Panels $a$ and $b$ for the analytical force-free field, Panels $c$ and $d$ for the modeled solar-like force-free field, and Panels $e$ and $f$ for the observed non-force-free field. Here the left panels are for the $F_x/F_p$ and $F_y/F_p$ and left panels are for the $F_z/F_p$. The lower x-axis is the added noise level of $\sigma_z^0$ and the upper x-axis is the added noise level of $\sigma_x^0$ or $\sigma_y^0$, all in the unit of Gauss. Since the noises we added are white noises, it would not change the total flux and hence MI. A calculation of the MI shows that the changes here are all less than 0.1\% so we do not plot it here.

In Figure 4, it can be seen that the measurement noises give little influence to $F_x/F_p$ and $F_y/F_p$ for the three representative fields. The variations of $F_x/F_p$ and $F_y/F_p$, with different noise levels, are all within 0.1 in magnitudes. However, the $F_z/F_p$ increases monotonously with the increase of the noise level. Panel $b$ shows that when $\sigma_z^0$ is larger than 10 G (100 G for $\sigma_x^0$ and $\sigma_y^0$), the $F_z/F_p$ has increased to a value larger than 0.1 for the analytical force-free field. For the modeled solar-like force-free field (Panel $d$), even when $\sigma_z^0$ = 5 G ($\sigma_x^0$ and $\sigma_y^0$ = 50 G), the $F_z/F_p$ has already increased to be over 0.1. These suggest that a truly force-free field may be estimated as non-force-free by the calculation of $F_z/F_p$.

For the non-force-free field (Panel $f$), the $F_z/F_p$ value also increases monotonically with the increase of noise level. This results in a situation that, when $\sigma_z^0$ is between 10 G and 15 G, the originally non-force-free field may appear as being force-free ($|F_x/F_p|\leq 0.1$, $\leq0.1$ and $|F_z/F_p|\leq0.1$). Note that a noise level between 10 G to 15 G in $\sigma_z^0$ (100 G to 150 G in $\sigma_x^0$ or $\sigma_y^0$) is right in the range of current measurement errors (70-200 G for transverse fields of HMI for example). So these plots deliver a serious warning of using $F_z/F_p$ to judge the force-freeness in the presence of measurement noises.

Figure 5 shows a further study on the influence of the noise level. Here we have added both the white noise and the sensitivity cutting. We first added the white noise at a level of $\sigma_z$ (again with $\sigma_x$ and $\sigma_y$ = 10 $\sigma_z$) and then omit the data points whose magnitudes are below 1$\sigma$ or 2$\sigma$. This is a situation more close to what is in the real observation: having the influence from the instrument sensitivity and measurement noise at the same time.

Plotted in the right panels of Figure 5 are the variations of $F_z/F_p$ with different noise levels, again the lower x-axis is the added noise level of $\sigma_z^0$ and the upper x-axis is the added noise level of $\sigma_x^0$ or $\sigma_y^0$, all in the unit of Gauss. The black lines show the same result as those in Figure 4, that is, without considering the influence of sensitivity. The red lines show the results of cutting at 1$\sigma$ level, and the blue lines for cutting at 2$\sigma$ level. Similarly as in Figure 3, the values of MI and NP are plotted in the right panels, with the same color-coding for the sensitivity cutting levels.

From the left panels of Figure 5, we see that the addition of the influence of instrument sensitivity upon the influence of white noise does not change the results too much from what we have already seen in Figure 4: Still a true force-free field may be estimated as being non-force-free and a non-force-free field may be estimated as being force-free if the noise level is high enough. Note that the changing point, where a wrong judgment may be made, is high only compared to those without any measurement noise. These changing points of noise level are actually right within the range of current measurement noise levels.

Taken the non-force-free fields as an example, when $\sigma_x^0$ or $\sigma_y^0$ is larger than 130G, in the case of 2$\sigma$ cutting, even though 95.6\% data points have been omitted, the noise in the remaining data points can still lead to a wrong estimation.

These result call for a serious caution on interpreting the $F_z/F_p$ measures when using observed magnetograms. Even cutting at a 2$\sigma$ level, the noises in the remaining 10\% data points can still affect the force-freeness estimation. In particular, note that most previous studies (Metcalf et al. 1995; Moon et al. 2002; Tiwari 2012; Liu et al. 2013) show that most active regions have their $F_z/F_p$ magnitudes larger than those of $F_x/F_p$ and $F_y/F_p$ and the judgment of force-free nature mainly depends on their $F_z/F_p$ magnitudes.

\section{Discussions}

In this section, we discuss on why the $F_z/F_p$ increases monotonously with the increase of noise level as shown in Figures 4 and 5. Our explanation also leads to an interesting judgment of the nature of force-freeness on the photosphere.

This monotonous increase with the noise level is actually buried in the form of Equation (4). In real observations where measurement noise is unavoidable, the vector magnetogram we obtained, that is, {\bf B}$^{\prime}=(B_x^{\prime},B_y^{\prime},B_z^{\prime})$, is actually ($B_x+\sigma_x$, $B_y+\sigma_y$, $B_z+\sigma_z$), where {\bf B}=(B$_x$,B$_y$,B$_z$) denotes the true field and ($\sigma_x$, $\sigma_y$, $\sigma_z$) are corresponding noise levels. So, applying Equation (4) to {\bf B}$^{\prime}$, what we actually calculated are:
\begin{equation}
\begin{aligned}
F_x^{\prime} & = -\frac{1}{4\pi} \int [(B_x +\sigma_x)( B_z+\sigma_z)]\, dx\,dy\\
&=-\frac{1}{4\pi} \int (B_x  B_z+ B_x \sigma_z + B_z \sigma_x + \sigma_x \sigma_z)\, dx\,dy\,, \\
F_y^{\prime} & = -\frac{1}{4\pi} \int [(B_y +\sigma_y)( B_z+\sigma_z)]\, dx\,dy \\
&=-\frac{1}{4\pi} \int (B_y  B_z+ B_y \sigma_z + B_z \sigma_y + \sigma_y \sigma_z)\, dx\,dy\,, \\
F_z^{\prime} & = \frac{1}{8\pi} \int[ ( B_t+\sigma_t)^2\,-(B_z +\sigma_z)^2\,] \,dx\,dy\\
 &= \frac{1}{8\pi} \int(B_t^2 +2B_t \sigma_t + \sigma_t^2\,-B_z^2 -2B_z \sigma_z - \sigma_z^2) \,dx\,dy\,, \\
\end{aligned}
\end{equation}
\noindent where $B_t$ is the horizontal field ($B_t^2=B_x^2+B_y^2$) and $\sigma_t$ is the horizontal field noise ($\sigma_t=\sqrt{\sigma_x^2+\sigma_y^2}$).

In this Equation, the first-order terms, such as $B_x \sigma_z$, $B_y \sigma_z$ and $B_t \sigma_t$, will cancel with each other by the integration due to the fact that the added noise is a white noise. However, the second-order terms, that is, $\sigma_t^2-\sigma_z^2$, would not cancel with each other by the integration and will accumulate due to the fact that $\sigma_t >> \sigma_z$. This is why the changes of $F_x/F_p$ and $F_y/F_p$ is almost negligible with the noise increasing, whereas $F_z/F_p$ increases dramatically with the noise increasing, as shown in Figures 4 and 5.

We noticed that Moon et al. (2002) also estimated the resulting error of $F_z/F_p$ caused by measurement noise. However, they only considered the first-order term whose contribution is indeed small. Metcalf et al. (1995) realized this problem by stating that ``since the $F_z$ and $F_0$ integrals use the square of the field strengths, the noise in the measurements will be magnified in the results". However, the approach they took is only to `cut' at a high level as of $150 G$, an approach that we have demonstrated that would not remove the influence of the noises as shown in Figure 5. Our analysis has shown that, even cutting at 2$\sigma$  level, the noises in the remaining even 10\% data points will significantly influence the estimation of $F_z/F_p$. To reduce this effect, we suggest that more accurate measurements are necessary, better to control the noise level less than 40 G for the transverse fields.

Having shown that the noises will increase the value of $F_z/F_p$, it implies that previous studies, based on observed magnetograms whose data contain noises, might have overestimated the value of $F_z/F_p$. So, if the measured $F_z/F_p$ is already negative, the true $F_z/F_p$ may have an even more negative value. We noticed that the $F_z/F_p$ values are negative for almost all the active regions in Moon et al. (2002) and sunspots in Tiwari (2012), so we estimate that the true $F_z/F_p$ values may have even more negative values and the true magnitudes of $|F_z/F_p|$ may be larger than their current estimates. This implies that the photospheric magnetic fields may be non-force-free as Metcalf et al. (1995) and Liu et al. (2013) have stated, rather than `close to be force-free' as Moon et al. (2002) and Tiwari (2012) have stated.

\section{Summary}

We studied in this investigation on how the measurement issues could influence the estimation and judgement of the force-freeness of the magnetic fields. We used three representative fields, analytical force-free solutions, modeled solar-like force-free fields and observed non-force-free fields, and mimic the effects of different field of view, instrument sensitivity and measurement noise to an extent similar as to what current photospheric measurements suggest.

We find that the measurement issues can bring certain effects on the results of force-freeness estimation. Among these factors, the problems of field of view and instrument sensitivity would not significantly influence the force-free measures if the vertical magnetic flux imbalance is less than 10\%. However, the measurement error (white noise) gives a significant impact. It may make a true force-free field be estimated as being non-force-free and a non-force-free field be estimated as being force-free.

This is because the $F_z$ and $F_p$ integrals in Low (1985)'s formula use the square of the field strengths, the noises in the measurement will be magnified by the integration instead of canceling with each other. Cutting the magnetogram at a high sensitivity level would not help. Our example shows that even cutting at 2$\sigma$ level, the noise in the remaining 10\% data points can still affect the force-freeness estimation. To decrease this effect, the noise level of the measurement needs to be controlled at a level of less than 40 G for the transverse fields.

Taking into account of current measurement noise levels, our results suggest that caution should be taken when using the observed magnetograms and Low's formula to estimate the force-freeness and make a judgment. Our analysis also indicates that the true photospheric magnetic fields might be non-force-free as Metcalf et al. (1995) and Liu et al. (2013) have suggested, rather than `close to be force-free' as Moon et al. (2002) and Tiwari (2012) have estimated.

A further note to put here is that different parts of the sunspots, such as the umbra, inner penumbra and outer penumbra, may have different properties of the force-freeness. For example, Tiwari et al. (2012) found that the umbral fields are more force-free than the penumbra fields and the inner penumbra are more force-free than the middle and outer penumbra. However, this result was obtained by estimating the vertical tension force in different parts, a method not requiring magnetic flux balance (though requiring an estimate of the plasma density which is a tough task itself). So it would be difficult to use the method described in this paper to check this statement, although it is an interesting phenomena that deserves further investigations.

\acknowledgements

We thank the anonymous referee for helpful comments and suggestions that improved the paper. We acknowledge the support of the National Natural Science Foundation of China (Grants No. 11125314 and No. U1531247) and the Strategic Priority Research Program (Grant No. XDB09000000) of the Chinese Academy of Sciences.


\newpage

\begin{figure}[!ht]
\centerline{\includegraphics[width=0.95\textwidth]{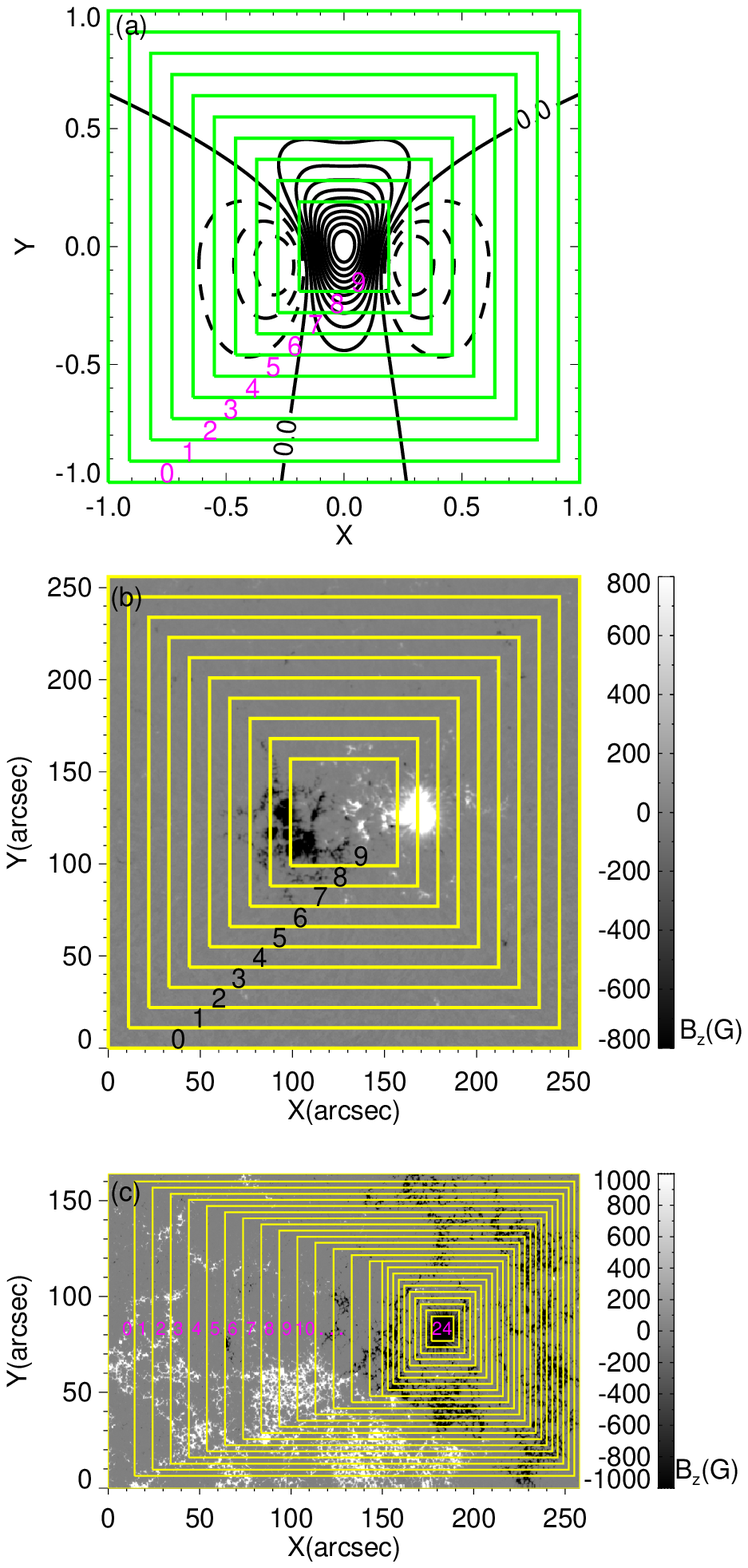}}
\caption{Field configurations of the three representative fields. ($a$) The analytical force-free field: contours show the vertical field strength with solid contours indicating $B_z>0$ field and dashed contours showing $B_z<0$ field, the contour interval is 180 G. ($b$) The modeled solar-like force-free field: $B_z$ map of the extrapolated field based on NOAA 11072 magnetogram observed by HMI$/$SDO at 05:00 UT on 23 May 2010, . ($c$) The non-force-free field: B$_z$ map of NOAA 10960 as observed, obtained by SP$/$Hinode on 2007 June 7. The green or yellow rectangles in the three panels mimic different FOVs to be studied in Section 3. See text for more details.}
\end{figure}

\begin{figure}[!ht,placement specifier]
\centerline{\includegraphics[width=0.95\textwidth]{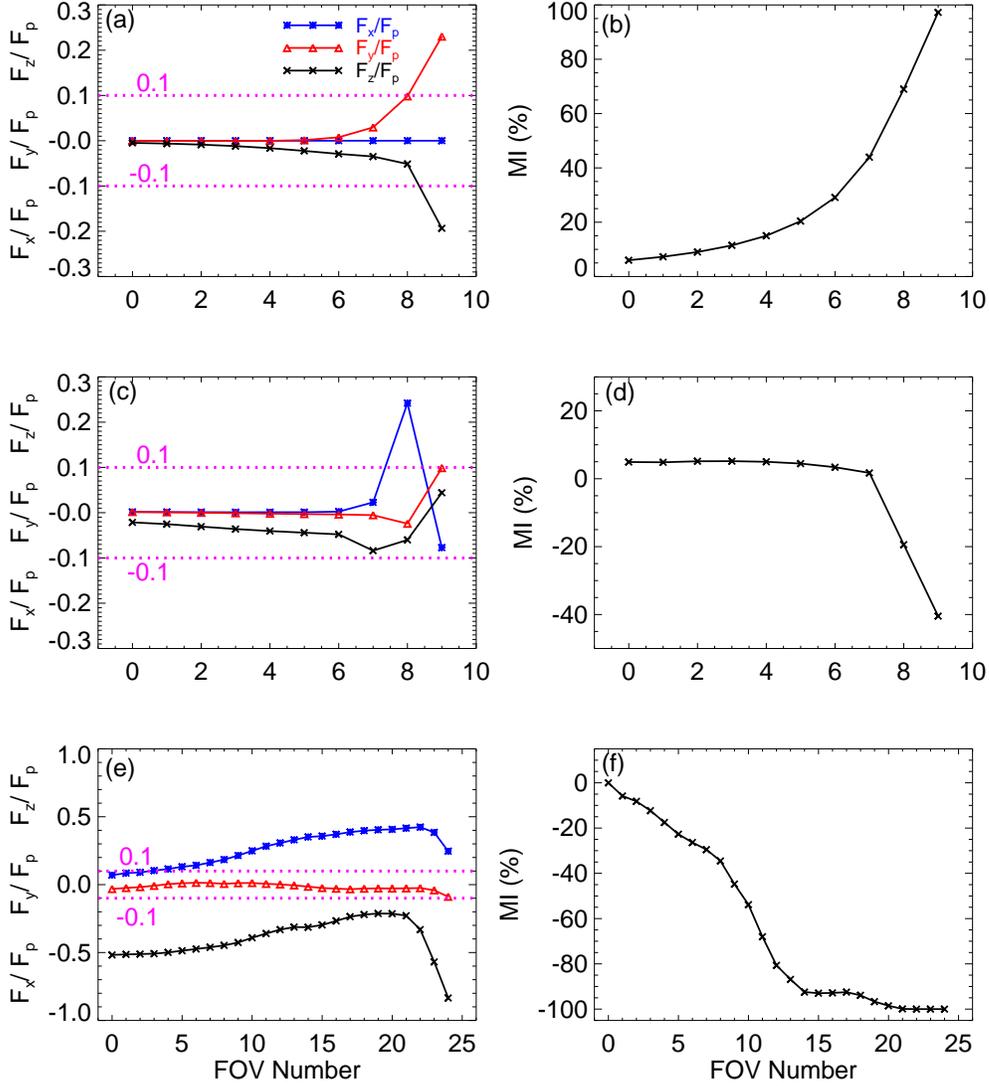}}
\caption{Influence of different FOVs on estimating the force-freeness. ($a$): The variation of $F_x/F_p$ (blue line), $F_y/F_p$ (red line) and $F_z/F_p$ (black line) for the analytical force-free field; ($b$): The variation of MI for the analytical force-free field; ($c,d$): Same as in Panels $a$ and $b$, but for the modeled solar-like force-free field; ($e,f$): Same as in Panels $a$ and $b$, but for the observed non-force-free field. }
\end{figure}

\begin{figure}[!ht]
\centerline{\includegraphics[width=0.95\textwidth]{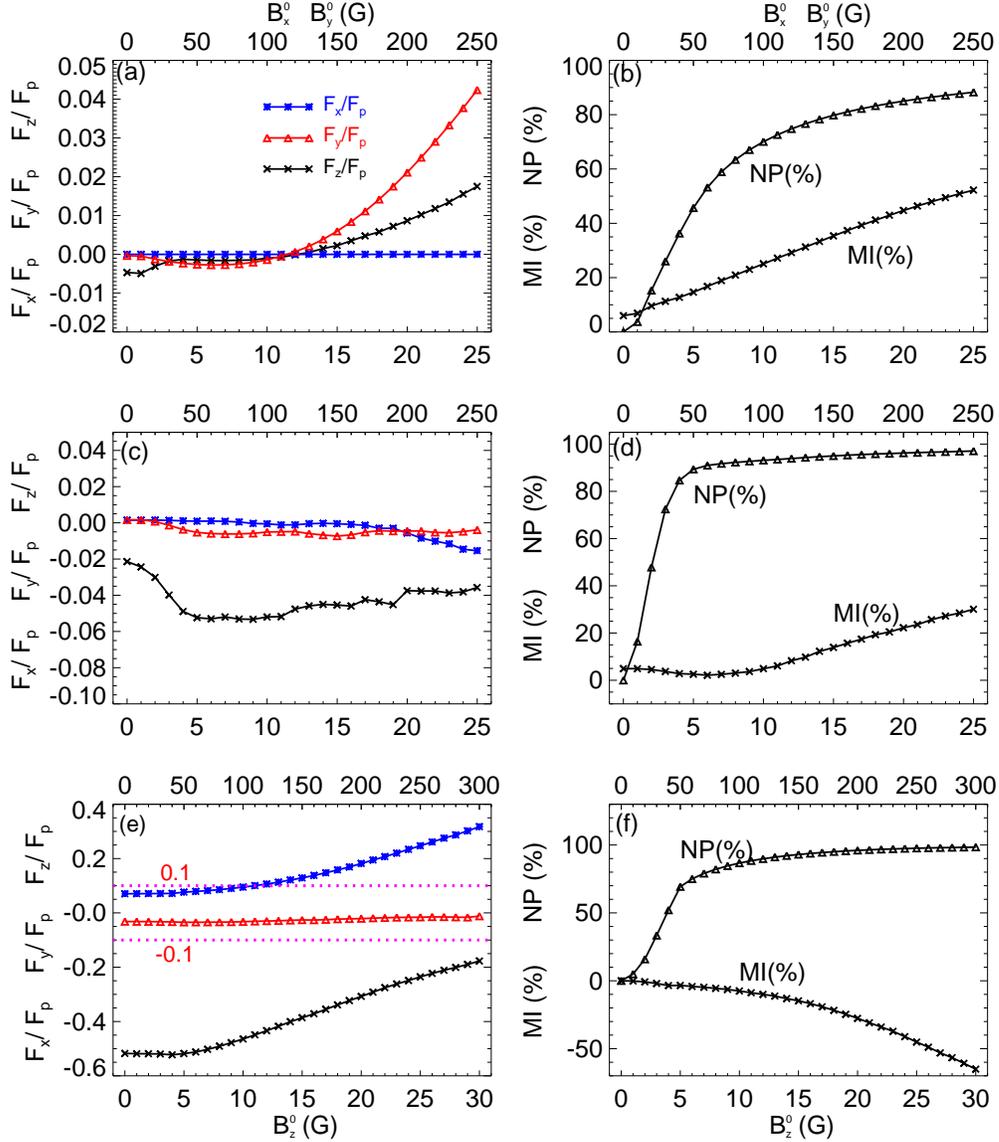}}
\caption{Influence of the instrument sensitivity on estimating the force-freeness. ($a$): The variation of $F_x/F_p$ (blue line), $F_y/F_p$ (red line) and $F_z/F_p$ (black line) for the analytical force-free field; ($b$): The variation of MI (with cross symbols) and NP (with triangle symbols) for the analytical force-free field; ($c,d$): Same as in Panels $a$ and $b$, but for the modeled solar-like force-free field; ($e,f$): Same as in Panels $a$ and $b$, but for the non-force-free field. The lower x-axis is the $B_z^0$ and the upper x-axis is the $B_x^0$ or $B_y^0$, both in the unit of Gauss.}
\end{figure}

\begin{figure}[!ht]
\centerline{\includegraphics[width=0.95\textwidth]{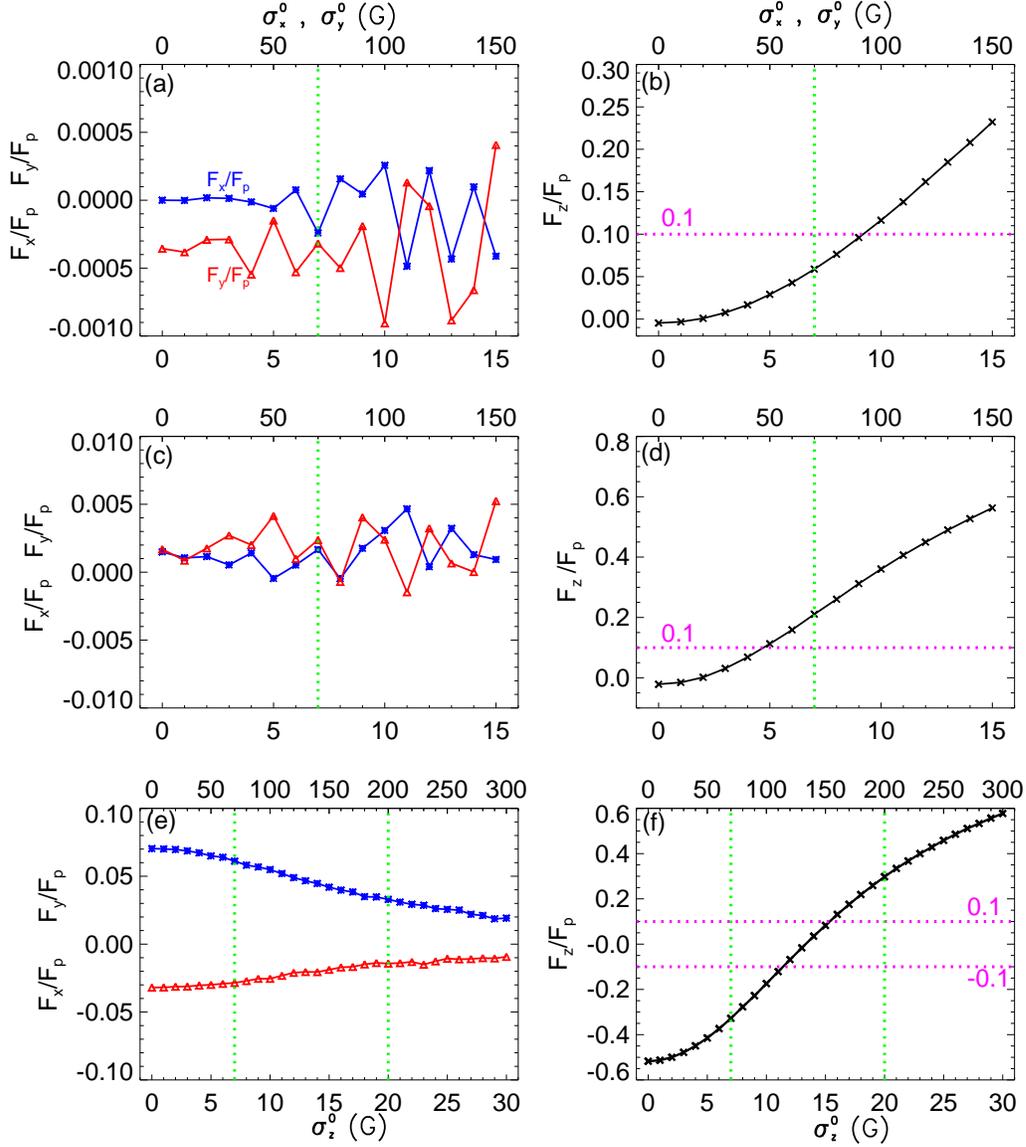}}
\caption{Influence of the measurement noise on estimating the force-freeness. ($a$): The variations of $F_x/F_p$ (blue line) and $F_y/F_p$ (red line) with different noise levels for the analytical force-free field; ($b$): The variation of $F_z/F_p$ (black line) for the analytical force-free field; ($c,d$): Same as in Panels $a$ and $b$, but for the modeled solar-like force-free field; ($e,f$): Same as in Panels $a$ and $b$, but for the non-force-free field. The lower x-axis is the $\sigma_z^0$ and the upper x-axis is the $\sigma_x^0$ or $\sigma_y^0$, both in the unit of Gauss. The vertical green dashed lines show the range of current measurement noise levels (70 - 200 G for transverse fields).}
\end{figure}

\begin{figure}[!ht]
\centerline{\includegraphics[width=0.95\textwidth]{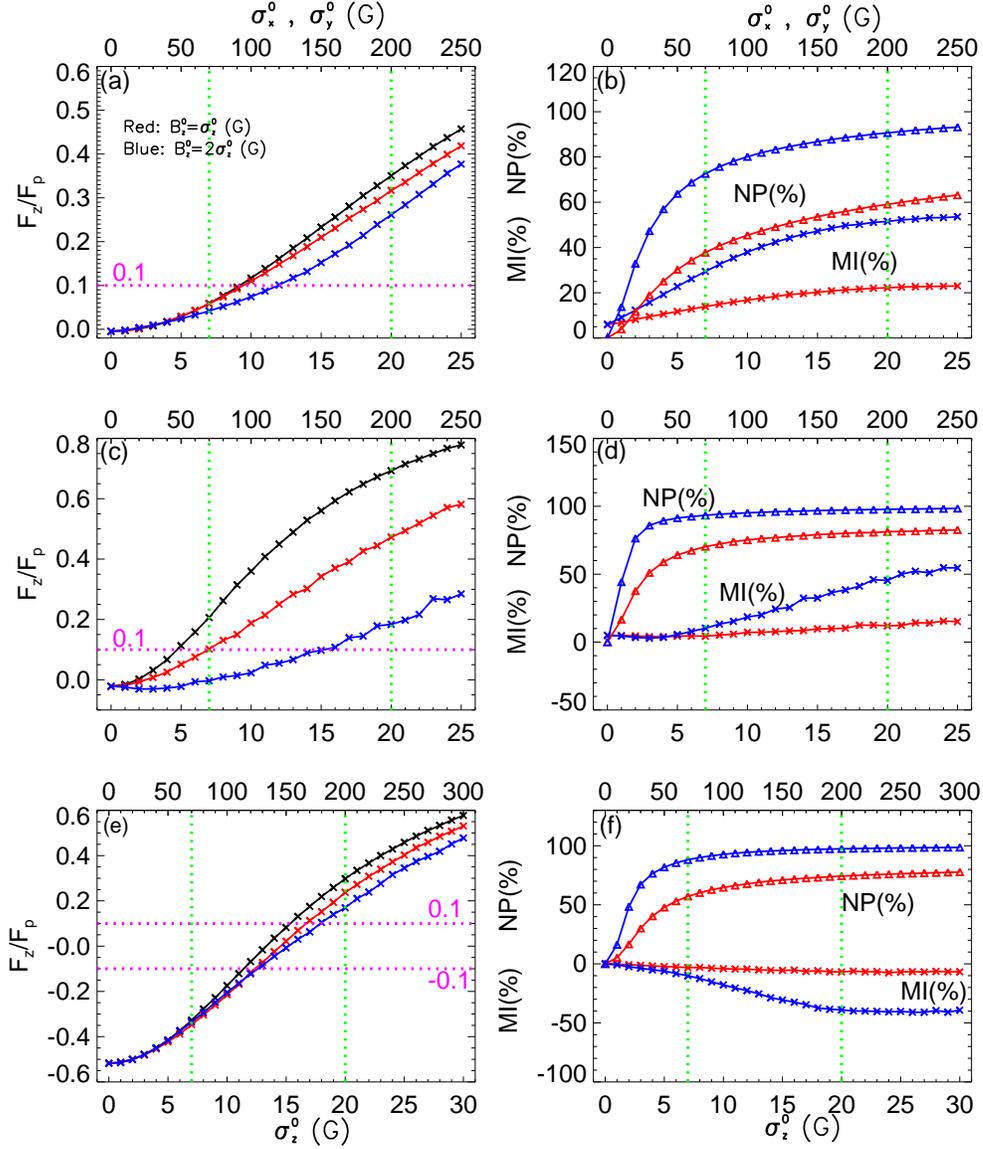}}
\caption{Influence of the measurement noise and sensitivity on estimating the force-freeness. ($a$): The variation of $F_z/F_p$ for the analytical force-free field, black line: same as in Figure 4$b$, adding noise without sensitivity cutting, red line: adding noise with 1$\sigma$ cutting, blue line: adding noise with 2$\sigma$ cutting; ($b$): The variation of MI and NP for the analytical force-free field, with the same color-coding as in Panel a; ($c,d$): Same as in Panels $a$ and $b$, but for the modeled solar-like force-free field; ($e,f$): Same as in Panels $a$ and $b$, but for the non-force-free field. The lower x-axis is the $\sigma_z^0$ and the upper x-axis is the $\sigma_x^0$ or $\sigma_y^0$, both in the unit of Gauss. The vertical green dashed lines show the range of current measurement noise levels (70 - 200 G for transverse fields).}
\end{figure}

\end{document}